\documentclass[prc,nofootinbib,showpacs]{revtex4}

\usepackage{graphicx}
\usepackage[usenames,dvipsnames]{color}
\usepackage{amsmath,amssymb,bbold}
\usepackage{hyperref}
\usepackage{comment}

\begin{document}

\title{Electrical conductivity and Hall conductivity of hot and dense hadron gas in a magnetic field: a relaxation time approach}

\author{Arpan Das}
\email{arpan@prl.res.in}
\author{Hiranmaya Mishra}
\email{hm@prl.res.in}
\affiliation{Theory Division, Physical Research Laboratory,
Navrangpura, Ahmedabad 380009, India}
\author{Ranjita K. Mohapatra}
\email{ranjita@iitb.ac.in}
\affiliation{Department of Physics, Indian Institute of Technology Bombay, Mumbai, 400076, India}

\begin{abstract}
We estimate the electrical conductivity and the Hall conductivity of hot and dense hadron gas using the relaxation time approximation of 
the Boltzmann transport equation in the presence of electromagnetic field. We have investigated the temperature and the baryon chemical
potential dependence of these transport coefficients in presence of magnetic field. The explicit calculation is performed within the ambit
of the hadron resonance gas model. We find that the electrical conductivity decreases in the presence of magnetic field. The Hall conductivity 
on the other hand shows a non monotonic behavior with respect to the dependence on magnetic field. We argue that for a pair plasma (particle-anti 
particle plasma) where $\mu_B=0$,  Hall conductivity vanishes. Only for non vanishing  baryon chemical potential 
Hall conductivity has non zero value. We also  estimate the electrical conductivity and the Hall conductivity as a function of the center
of mass energy along the freeze out curve as may be relevant for relativistic heavy ion collision experiments.
\end{abstract}

\pacs{25.75.-q, 12.38.Mh}
\maketitle

\section{INTRODUCTION}

\label{intro}

  Transport coefficients of strongly interacting matter created in the relativistic heavy ion collision
experiments are of great importance for a comprehensive 
understanding of the hot and dense QCD (quantum chromodynamics) medium produced in these experiments.
Experimental data and theoretical models 
give a strong hint for the formation of quark-gluon plasma (QGP) in the initial stage of heavy ion collisions and its subsequent hadronization, at relativistic heavy ion collider (RHIC) and the large hadron collider (LHC). In the dissipative relativistic hydrodynamical model of the hot and dense medium as well as for
transport simulations, which are being used to describe the evolution of the strongly interacting matter in heavy ion collision, transport coefficients, e.g. shear and bulk viscosity etc plays an important role. In fact, it has been shown that a small shear viscosity to entropy ratio ($\eta/s$) is necessary to explain the flow data, in the context of hydrodynamical modeling for the evolution of the QGP medium subsequent to a heavy ion collision.\cite{HeinzSnellings2013,RomatschkeRomatschke,KSS}. 
The smallness of $\eta/s$ and its connection with the lower bound for the same, $\eta/s=\frac{1}{4\pi}$ using 
gauge gravity duality (AdS/CFT correspondence) has motivated  a large number of investigations in understanding the behaviour of transport coefficients
from a microscopic  theory \cite{KSS}. The bulk viscosity $\zeta$, also plays a significant role  
in the dissipative hydrodynamics describing the QGP evolution \cite{gavin1985,kajantie1985, DobadoTorres2012,sasakiRedlich2009,sasakiRedlich2010,KarschKharzeevTuchin2008,
FinazzoRougemont2015,WiranataPrakash2009,JeonYaffe1996}. The bulk viscosity encodes the conformal measure $(\epsilon-3P)/T^4$ of the system and lattice QCD simulations shows a non monotonic
behaviour of both $\eta/s$ and $\zeta/s$  near the  critical temperature $T_c$.  \cite{DobadoTorres2012,sasakiRedlich2009,sasakiRedlich2010,KarschKharzeevTuchin2008,
FinazzoRougemont2015,WiranataPrakash2009,JeonYaffe1996}.
In case of non central heavy ion collisions, due to the collision geometry, a large magnetic field is also expected to be produced. The magnitude of the produced magnetic field at the initial stages in these collisions are expected to be rather large, at least of the order of  several $m_{\pi}^2$ \cite{mclerran2008,skokov}. Since the strength
of the magnetic field is of hadronic scale, the effect of the magnetic field on the QCD medium can be significant. This prompts  a deeper understanding of the QGP, as well as the subsequent hadronic medium in magnetic field. The strong magnetic field produced in relativistic heavy ion collision experiments along with the presence of deconfined non abelian QCD matter bring some exciting possibilities specifically different CP violating effects such as chiral magnetic effect
and chiral vortical effect  \cite{kharzeevbook}. Magnetohydrodynamic simulations which incorporate the large scale behavior of the thermal medium in the presence of a dynamical electromagnetic field in a self-consistent manner, have been used to study the flow coefficient of the of the strongly interacting matter \cite{MHD1,MHDajit}.
Phenomenological manifestation of magnetic field requires that the magnetic field should survive for at least few Fermi proper time. Evolution of the electromagnetic field in a conducting plasma is intimately related to the electrical 
conductivity, $\sigma^{el}$ of the medium \cite{MHD1,MHDajit,TuchinMHD,MoritzGreif,electricalcond1,electricalcond2,electricalcond3,electricalcond4,
electricalcond5,electricalcond6,electricalcond7,electricalcond8,electricalcond9,electricalcond10,electricalcond11,electricalcond12,
electricalcond13,electricalcond14,electricalcond15}.  Different approaches like perturbative QCD, and different effective models etc have been used to estimate various transport coefficients for the QCD matter  \cite{danicol2018,PrakashVenu,WiranataPrakash2012,
KapustaChakraborty2011,Toneev2010,Plumari2012,Gorenstein2008,Greiner2012,TiwariSrivastava2012,GhoshMajumder2013,Weise2015,GhoshSarkar2014,
WiranataKoch,WiranataPrakashChakrabarty2012,Wahba2010,Greiner2009,KadamHM2015,Kadam2015,Ghoshijmp2014,Demir2014,Ghosh2014,smash,bamps,bamps2,
urqmd1,GURUHM2015, ranjitahm,amanhm1,amanhm2,arpanhm}. The other transport coefficient that 
plays a significant role in the hydrodynamical evolution at non zero baryon densities is the thermal conductivity. \cite{danicol2014,Kapusta2012}.

  In the present work, we investigate the electrical and the Hall conductivity of the hot and dense hadron gas produced in the subsequent evolution of QGP, in heavy-ion collisions. The Hall effect is the production of an induced electric current transverse to the electric field and to an applied magnetic field (perpendicular to the electric field), in a conducting medium. Hall effect is a manifestation of the diffusion of electric charge perpendicular to the electric field. Necessary requirements of Hall current are the presence of mobile charge carries, an external electric and magnetic field perpendicular to each other. Electric field and the magnetic field produced in heavy ion collisions can have such configurations \cite{tuchin1,tuchin2}, and, therefore, it is natural to investigate the Hall effect in the strongly interacting matter. Indeed in the presence of magnetic field the Hall conductivity for the QGP medium has been recently investigated in Ref.\cite{feng2017}.

  In the present investigation, we study the electrical and Hall conductivity for the hot and dense hadronic matter in a magnetic field using the hadron resonance gas model (HRGM) within the framework of relaxation time approximation (RTA). It may be noted that in the usual condensed matter systems e.g. semiconductors, the Hall effect requires applied electric and magnetic field perpendicular to each other and in these cases either the electrons or the holes are the dominant charge carriers \cite{semiconductor}. 
  Further in the context of electron-ion plasma the mobility of the positive and negative charge carriers are different, which can give rise to net Hall current. On the other hand, for pair plasma (e.g. electron positron plasma) due to the opposite gyration of the positive and negative charge carriers the net Hall current vanishes \cite{pairplasma1,pairplasma2,pairplasma3}. This will be similar in quark-gluon plasma (QGP) with zero baryon chemical potential. However,  the situation is different at finite baryon chemical potential, when the number of baryons and anti-baryons are different. At finite baryon chemical potential unequal numbers of positive and negative charge carriers can generate a net Hall current. At finite baryon chemical potential only the net baryons contributes to the Hall current and the mesonic contribution to the Hall current exactly cancels with the opposite Hall current due to its antiparticle . A baryon-rich strongly
 interacting medium is expected to create at heavy-ion collisions experiments at Facility for Antiproton and Ion Research (FAIR) at Darmstadt \cite{fairref} and in 
Nuclotron-based Ion Collider fAcility (NICA) at Dubna \cite{nica}. In these cases, a thermalized strongly interacting medium is expected to produce at finite baryon density, which is not electrically charged neutral. 
Keeping the above motivation in mind, we calculate the electrical conductivity and Hall conductivity of hadron resonance gas in a magnetic field within the kinetic theory
framework. Let us note that the large magnetic field produced in heavy ion collision can be sustained in the QGP medium with finite electrical conductivity. One would expect however at later times when quark hadron transition takes place during the course of evolution, the magnetic field may not be as strong as in the QGP phase. In the present approach therefore we attempted to estimate the transport coefficients where the phase space and the single particle energies are not affected by magnetic field through Landau quantization. On the other hand, the effect of magnetic field in the medium enters through the cyclotron frequency of the individual hadrons.

 The hadron resonance gas (HRG) model describes, the hadronic phase of the strongly interacting medium created in heavy ion collisions, at chemical freeze-out \cite{HRG1,HRG2}. This model has been extensively and successfully used to reproduce the experimental result 
of the thermal abundance of different particle ratios in the heavy ion collisions 
\cite{HRG3}. In the simplest case when the strange and non strange particles freeze out in the same manner, HRG model has only two parameters temperature ($T$) and baryon chemical potential ($\mu_B$). It has been shown that 
 in the presence of narrow-resonances, the thermodynamics of interacting gas of hadrons can be approximated by the non-interacting
gas of hadrons and its resonances \cite{HRG4,HRG5}. Because of this phenomenological fact, thermodynamics of strongly interacting hadronic system can be easily studied using non interacting hadrons and its resonances, where the interactions between different hadrons are encoded as the resonances. 
Due to this simplified non interacting quasi particle description, HRG model has been well explored regarding
thermodynamics \cite{thermodynamicsHRG1,thermodynamicsHRG2}, conserved charge fluctuations\cite{hrgfluc3,hrgfluc4,hrgfluc5,hrgfluc6,
hrgfluc7} as well as transport coefficients for
hadronic matter\cite{MoritzGreif,PrakashVenu,WiranataPrakash2012,
KapustaChakraborty2011,Toneev2010,Plumari2012,Gorenstein2008,Greiner2012,TiwariSrivastava2012,GhoshMajumder2013,Weise2015,GhoshSarkar2014,
WiranataKoch,WiranataPrakashChakrabarty2012,Wahba2010,Greiner2009,KadamHM2015,Kadam2015,Ghoshijmp2014,Demir2014,Ghosh2014,smash,
bamps,bamps2,electricalcond2,electricalcond3,Plumari2012,urqmd1,GURUHM2015}. Some improvements has also been done on the ideal HRG model e.g. including excluded volume HRG model \cite{stockerRischke,GURUHM2015}. 
We would like to mention here that, although the electrical conductivity has been well explored in HRG model, the Hall conductivity and the effect of magnetic field on electrical and Hall conductivity has not been studied earlier for hadronic system. The present investigation is a first step in that direction.

This paper is organized as follows, in Sec. \ref{formalism} we introduce the formalism to estimate electrical conductivity and Hall conductivity using kinetic theory within relaxation
time approximation.  In Sec. \ref{HRGmodel} we briefly discuss the HRG model and estimate the relaxation time within the same model.
In Sec. \ref{results} we present and discuss the results for electrical and Hall conductivity. Finally we summarize our work with an outlook in the
conclusion section.

\section{Boltzmann equation in relaxation time approximation}
\label{formalism}
The relativistic Boltzmann transport equation (RBTE) of a charged particle of single species in the presence of external electromagnetic field can be written as \cite{feng2017}, 

\begin{align}
 p^{\mu}\partial_{\mu}f(x,p)+eF^{\mu\nu}p_{\nu}\frac{\partial f(x,p)}{\partial p^{\mu}} = \mathcal{C}[f],
 \label{equ1}
\end{align}
where $F^{\mu\nu}$ is the electromagnetic field strength tensor, $e$ is the electric charge of the particle and $\mathcal{C}[f]$ is the collision integral. In the relaxation time
approximation (RTA) the collision integral can be written as, 

\begin{align}
 \mathcal{C}[f]\simeq -\frac{p^{\mu}u_{\mu}}{\tau}(f-f_0)\equiv -\frac{p^{\mu}u_{\mu}}{\tau}\delta f ,
 \label{equ2}
\end{align}
where, $u_{\mu}$ is the fluid four velocity that in the local rest frame in $(1,\vec{0})$ and $\tau$ is the relaxation time which determine the time scale for the system to relax towards the equilibrium state characterized by the distribution function $f_0$. 
In the relaxation time approximation the underlying assumption is the system is slightly away from equilibrium and then it relaxes towards 
equilibrium with the time scale $\tau$. In relaxation time approximation the external electromagnetic field can not be very large with respect to the 
characteristic scale of the thermal system, hence we are not considering the Landau quantization of the charged particles in a magnetic field. This assumption is plausible in the context of hadronic medium where the strength of the initial magnetic field is relatively smaller with respect to the initial large magnetic field produced in heavy ion collisions.  In this work we have considered Boltzmann distribution function, which has been widely used for the Hadron 
resonance gas model. The equilibrium distribution function satisfies, 

\begin{align}
 \frac{\partial f_0}{\partial \vec{p}}=\vec{v}\frac{\partial f_0}{\partial \epsilon}, ~~ \frac{\partial f_0}{\partial \epsilon}=-\beta f_0,
 ~~f_0 = g e^{-\beta(\epsilon\pm\mu_B)}
 \label{equ3}
\end{align}
 where the single particle energy is $\epsilon(p)=\sqrt{\vec{p}^2+m^2}$, $g$ is the degeneracy, $\mu_B$ is the baryon chemical potential and $\beta=1/T$, is the 
 inverse of temperature. Using Eq.\eqref{equ2}, the Boltzmann equation \eqref{equ1} can be recasted as,
 
 \begin{align}
  \frac{\partial f}{\partial t}+\vec{v}.\frac{\partial f}{\partial\vec{r}}+e\bigg[\vec{E}+\vec{v}\times\vec{B}\bigg]
  .\frac{\partial f}{\partial\vec{p}}=-\nu(f-f_0),
  \label{equ4}
 \end{align}
 where $\nu=1/\tau$. In case of uniform and static  medium both $f$ and $f_0$ are independent of the time and space \cite{feng2017}.  In such a case Eq.\eqref{equ4} becomes, 
 
 \begin{align}
  -e\bigg[\vec{E}+\vec{v}\times\vec{B}\bigg]\frac{\partial f}{\partial\vec{p}} & = \nu(f-f_0),
  \label{equ5}
 \end{align}
 Without loss of generality, we take $\vec{E}=E\hat{x}$ and $\vec{B}=B\hat{z}$. For this representation of electric field 
and magnetic field, Eq.\eqref{equ5} can be rearranged as, 

\begin{align}
 \bigg(\nu-eB\bigg(v_x\frac{\partial}{\partial p_y}-v_y\frac{\partial}{\partial p_x}\bigg)\bigg)f(p)=\nu f_0(p)
 -eE\frac{\partial}{\partial p_x}f_0(p). 
 \label{equ6}
\end{align}

In order to solve Eq.\eqref{equ6}, we take the following ansatz of the distribution function $f(p)$ \cite{feng2017},

\begin{align}
 f(p)=f_0-\frac{1}{\nu}e\vec{E}.\frac{\partial f_0(p)}{\partial \vec{p}}-\vec{\Xi}.\frac{\partial f_0(p)}{\partial \vec{p}}.
 \label{equ7}
\end{align}

Using the ansatz given in Eq.\eqref{equ7}, we can simplify the Eq.\eqref{equ6}, 

\begin{align}
 \frac{eB}{\nu}eE\frac{v_y}{p}-eB\bigg(v_x\Xi_y-v_y\Xi_x\bigg)\frac{1}{p}+\nu\bigg(\Xi_x\frac{p_x}{p}+\Xi_y\frac{p_y}{p}
 +\Xi_z\frac{p_z}{p}\bigg)=0.
 \label{equ9}
\end{align}

Eq.\eqref{equ9} should be satisfied for any value of the velocity, hence we get $\Xi_z=0$. Comparing the coefficient of $v_y$ and $v_x$ on both sides 
of Eq.\eqref{equ9} one obtains, 

\begin{align}
   \omega_c\left(\frac{eE}{\nu}\right)+\omega_c\Xi_x+\nu\Xi_y=0,
   \label{sol1}
 \end{align}
and, 
\begin{align}
   -\omega_c\Xi_y+\nu\Xi_x=0.
   \label{sol2}
\end{align}

In the above equation $\omega_c=eB/\epsilon(p)$, is the cyclotron frequency of the charged particle. Solving the Eq.\eqref{sol1} and Eq.\eqref{sol2} for $\Xi_x$ and $\Xi_y$ one obtains, 

\begin{align}
 \Xi_x=-eE\frac{\omega_c^2}{\nu(\nu^2+\omega_c^2)}, ~~~\Xi_y=-eE\frac{\omega_c}{\omega_c^2+\nu^2}.
 \label{equ12}
\end{align}

Using Eq.\eqref{equ12}, the ansatz for the distribution function $f(p)$ given in the Eq.\eqref{equ7} can be simplified as, with, $p=|\vec{p}|$ 

\begin{align}
 f(p) & =f_0-eE\frac{p_x}{p}\frac{\partial f_0}{\partial p}\frac{\nu}{\nu^2+\omega_c^2}
 +eE\frac{p_y}{p}\frac{\partial f_0}{\partial p}\frac{\omega_c}{\omega_c^2+\nu^2}\nonumber\\
 & =f_0-eEv_x\left(\frac{\partial f_0}{\partial \epsilon}\right)\frac{\nu}{\nu^2+\omega_c^2}
 +eEv_y\left(\frac{\partial f_0}{\partial\epsilon}\right)\frac{\omega_c^2}{\omega_c^2+\nu^2
 }
 \label{equ13}
\end{align}

Electric current is given by \cite{feng2017}, 

\begin{align}
 j^i=e\int\frac{d^3p}{(2\pi)^3}v^i\delta f =\sigma^{ij}E_j=\sigma^{el}\delta^{ij}E_j+\sigma^H\epsilon^{ij}E_j,
\end{align}
where $\epsilon_{ij}$ is the anti symmetric $2\times2$ unity tensor, with $\epsilon_{12}=-\epsilon_{21}=1$. Then the electrical and the Hall 
conductivity can be identified as, 

\begin{align}
 \sigma^{el}=e^2\int\frac{d^3p}{(2\pi)^3}v_x^2\left(-\frac{\partial f_0}{\partial \epsilon}\right)\frac{\nu}{\nu^2+\omega_c^2},
\end{align}

\begin{align}
 \sigma^{H}=e^2\int\frac{d^3p}{(2\pi)^3}v_y^2\left(-\frac{\partial f_0}{\partial \epsilon}\right)\frac{\omega_c}{\nu^2+\omega_c^2}.
\end{align}

Assuming an isotropic system the electrical conductivity and the Hall conductivity can be expressed as, 

\begin{align}
 \sigma^{el}=\frac{e^2}{3T}\int\frac{d^3p}{(2\pi)^3}\frac{p^2}{\epsilon^2} \frac{\nu}{\nu^2+\omega_c^2}f_0=
 \frac{e^2}{3T}\int\frac{d^3p}{(2\pi)^3}\frac{p^2}{\epsilon^2} \frac{1/\tau}{(1/\tau)^2+\omega_c^2}f_0,
 \label{equ17}
\end{align}

\begin{align}
 \sigma^{H}=\frac{e^2}{3T}\int\frac{d^3p}{(2\pi)^3}\frac{p^2}{\epsilon^2} \frac{\omega_c}{\nu^2+\omega_c^2}f_0=
 \frac{e^2}{3T}\int\frac{d^3p}{(2\pi)^3}\frac{p^2}{\epsilon^2} \frac{\omega_c}{(1/\tau)^2+\omega_c^2}f_0
 \label{equ18}
\end{align}

It is important to note that in the absence of magnetic field expression of the electrical conductivity as given in Eq.\eqref{equ17} exactly matches with the standard expression for the same obtained in relaxation time approximation \cite{electricalcond3,lataguruhm}. In this work we have used thermal averaged relaxation time.
Electrical and Hall conductivity of single species can be extended for a system consists of multiple charge particle species. In this case, 

\begin{align}
 \sigma^{el}=\sum_i
 \frac{e^2_i\tau_i}{3T}\int\frac{d^3p}{(2\pi)^3}\frac{p^2}{\epsilon^2_i} \frac{1}{1+(\omega_{ci}\tau_i)^2}f_0,
 \label{equ19}
\end{align}

\begin{align}
 \sigma^{H}= \sum_i\frac{e^2_i\tau_i}{3T}\int\frac{d^3p}{(2\pi)^3}\frac{p^2}{\epsilon^2_i} \frac{\omega_{ci}\tau_i}{1+(\omega_{ci}\tau_i)^2}f_0,
 \label{equ20}
\end{align}
where $e_i$, $\tau_i$ and $\omega_{ci}$ are electric charge, thermal averaged relaxation time and cyclotron frequency of the $i'$th charged particle species. From the Eq.\eqref{equ19} and  Eq.\eqref{equ20} it is clear that  in a pair plasma particle and anti particle contributes to the electrical conductivity is a same manner, but in the case of the Hall conductivity their behaviour is exactly opposite. Once the relaxation time $\tau$ and cyclotron frequency is known for each species, electrical conductivity and the Hall conductivity can be estimated using Eq.\eqref{equ19} and Eq.\eqref{equ20}. In this work we model the hadronic matter using the hadron resonance gas model and estimate relaxation time as well as conductivities.

\section{HADRON RESONANCE GAS MODEL}
\label{HRGmodel}
The thermodynamic potential of a gas of hadrons and resonances at finite temperature ($T$) and baryon chemical potential ($\mu_B$) can be written as \cite{KadamHM2015},

\begin{equation}
 \log Z(\beta,\mu_B,V)=\int dm (\rho_M(m)\log Z_b(m,V,\beta,\mu_B)+\rho_B(m)\log Z_f(m,V,\beta,\mu_B)),
 \label{eq43}
\end{equation}
where, $V$ is the volume and $T = 1/\beta$ is the temperature of non-interacting point like hadrons and their resonances. Total partition function is composed of the partition functions of free mesons ($Z_b$) and baryons ($Z_f$) with mass $m$.  
Moreover, $\rho_B$ and $\rho_m$ denotes the spectral functions of free bosons (mesons) and fermions (baryons) respectively. The hadron properties are encoded in the spectral
densities. Once the spectral density is specified, taking derivatives of the 
logarithm of the partition function as given in 
Eq.\eqref{eq43}, with respect to the thermodynamic parameters T, $\mu_B$ and the volume $V$, various thermodynamic quantities can be calculated. In this investigation we have considered the  HRG model by taking all the  hadrons and their resonances below
a certain mass cutoff $\Lambda$ to estimate the thermodynamic potential. This is achieved by taking the spectral density
$\rho_{B/M}(m)$ of the following form, 

\begin{equation}
 \rho_{B/M}(m)=\sum_i^{M_i<\Lambda}g_i\delta(m-M_i),
 \label{eq44}
\end{equation}
 In  Eq.\eqref{eq44}, $M_i$ and $g_i$ are
the masses and the  corresponding degeneracy of the known hadrons and their resonances respectively. It is important to mention that, HRG with discrete
particle spectrum is very appealing because of its simple structure, however 
it can  explain
lattice QCD data for trace anomaly only up to temperature $\sim 130 $ MeV \cite{HRGMuller}. Along with the discrete
particle spectrum if one also includes
Hagedron spectrum then lattice QCD data for QCD trace anomaly up to $T \sim$ 160 MeV \cite{HRGMuller} can be reproduced. For details of thermodynamics 
of HRG model, see e.g. Ref.\cite{HRG1}. 

The relaxation time of particle $a$ having momentum $p_a$ and energy $E_a$ is defined as \cite{lataguruhm,KapustaChakraborty2011},

\begin{align}
\tau_a^{-1}(E_a)=\sum_{b,c,d}\int\frac{d^3p_b}{(2\pi)^3}\frac{d^3p_c}{(2\pi)^3}\frac{d^3p_d}{(2\pi)^3}W(a,b\rightarrow c,d)f_b^0,
\end{align}
where $W(a,b\rightarrow c,d)$ is the transition rate which can be expressed in terms of the transition amplitude $\mathcal{M}$ in the following way,

\begin{align}
W(a,b\rightarrow c,d)=\frac{(2\pi)^4\delta(p_a+p_b-p_c-p_d)}{2E_a2E_b2E_c2E_d}|\mathcal{M}|^2.
\end{align}

In the center of mass frame, the relaxation time ($\tau_a$) or equivalently interaction frequency ($\omega_a$) can be written as, 

\begin{align}
\tau_a^{-1}(E_a)\equiv \omega_a(E_a)=\sum_b\int\frac{d^3p_b}{(2\pi)^3}\sigma_{ab}v_{ab}f_b^0.
\label{equ24}
\end{align}

Here $\sigma_{ab}$ is the total scattering cross section for the process, $a(p_a)+b(p_b)\rightarrow c(p_c)+d(p_d)$ and $v_{ab}$ is the relativistic relative velocity between particle $a$ and $b$,

\begin{align}
 v_{ab}=\frac{\sqrt{(p_a.p_b)^2-m_a^2m_b^2}}{E_aE_b}.
\end{align}

In this work we shall be considering energy averaged relaxation time. One can obtain the energy independent relaxation time $\tau^a$ by averaging the relaxation time $\tau^a(E_a)$ over the distribution function $f_a^0(E_a)$ \cite{KapustaChakraborty2011,paramitahm2016},

\begin{align}
 \tau_a^{~-1}=\frac{\int f_a^0\tau^{-1}_a(E_a)dE_a}{\int f^0_adE_a}
 \label{equ26}
\end{align}

Using Eq.\eqref{equ24} and Eq.\eqref{equ26}, the energy averaged relaxation time
$(\tau_a)$,  can be given as \cite{GURUHM2015},

\begin{equation}
 \tau_a^{~-1}=\sum_b n_b\langle\sigma_{ab}v_{ab}\rangle,
 \label{equ27}
\end{equation}
where $n_b$ and $\langle\sigma_{ab}v_{ab}\rangle$ represents number density and thermal averaged cross section respectively. The thermal averaged 
cross section for the scattering process $a(p_a)+b(p_b)\rightarrow c(p_c)+d(p_d)$ is given as \cite{gondologelmini},

\begin{align}
 \langle\sigma_{ab}v_{ab}\rangle = \frac{\int d^3p_ad^3p_b \sigma_{ab}v_{ab}f_a^{0}(p_a)f_a^{0}(p_b)}{\int d^3p_ad^3p_b f_a^{0}(p_a)f_a^{0}(p_b)}.
\end{align}
Assuming hard sphere (of radius $r_h$) scattering for the cross section  ($\sigma=4\pi r_h^2$) and Boltzmann distribution function $f_a^{0}(p_a) = e^{-(E_a\pm\mu_a)/T}$, the thermal averaged cross section becomes, 
\begin{align}
 \langle\sigma_{ab}v_{ab}\rangle = \frac{\sigma\int d^3p_ad^3p_b v_{ab}e^{-E_a/T}e^{-E_b/T}}{\int d^3p_ad^3p_b e^{-E_a/T}e^{-E_b/T}}.
\end{align}

Note that in the above equation the chemical potential dependence gets canceled from the numerator and denominator. This is a consequence of the Boltzmann
approximation for equilibrium distribution function. After changing the momentum integration in the above equation to center of mass energy variable
($\sqrt{s}$) we get, 

\begin{align}
 \int d^3p_ad^3p_b v_{ab}e^{-E_a/T}e^{-E_b/T} = 2\pi^2T\int ds\sqrt{s}(s-4m^2)K_1(\sqrt{s}/T),
\end{align}
and
\begin{align}
 \int d^3p_ad^3p_b e^{-E_a/T}e^{-E_b/T} = \left(4\pi m^2 T K_2(m/T)\right)^2.
\end{align}

Thus the thermal averaged cross section can be written as, 

\begin{align}
 \langle\sigma_{ab}v_{ab}\rangle =\frac{\sigma}{8m^4TK_2^2(m/T)}\int_{4m^2}^{\infty}ds \sqrt{s}(s-4m^2)K_1(\sqrt s /T).
\end{align}

Here $\sqrt{s}$ is the center of mass energy, $K_1$ and $K_2$ are modified Bessel function of first order and second order respectively. When the
particles are of different species then the above equation can be written as, 

\begin{equation}
 \langle\sigma_{ab}v_{ab}\rangle = \frac{\sigma}{8Tm_a^2m_b^2K_2(m_a/T)K_2(m_b/T)}\int_{(m_a+m_b)^2}^{\infty}ds\times \frac{[s-(m_a-m_b)^2]}{\sqrt{s}}
 \times [s-(m_a+m_b)^2]K_1(\sqrt{s}/T),
\end{equation}
where $\sigma = 4\pi r_h^2$ is the total scattering cross section for the hard sphere. It is important to mention that while the cross
section $\sigma$ is independent of both temperature and baryon chemical potential, thermal averaged cross section $\langle\sigma v\rangle$
 is in general can depend on temperature ($T$) and chemical potential $\mu_B$ arising from the distribution functions.
 However in Boltzmann approximation $\langle\sigma v\rangle$ is independent of $\mu_B$  \cite{gondologelmini}. After evaluating the thermal averaged relaxation time using Eq. \eqref{equ27} for each species we 
estimate the electrical conductivity and the Hall conductivity of the hot and dense hadron gas using 
Eq.\eqref{equ19} and Eq.\eqref{equ20} respectively.

\section{results and discussions}
\label{results}
As mentioned earlier, for the hadron resonance gas model, we consider here all the hadrons and their resonances up to an upper cutoff in mass $\Lambda$ which we take as $\Lambda =  2.6$ GeV as is listed in Ref.\cite{pdg}. For a detailed list of particles we refer to  Appendix A of Ref.\cite{kapustaAlbr}. The other parameter that enters in the estimation of relaxation time calculation are the radii of the hard spheres. We have considered an uniform radius of
$r_h=0.5$ fm for all the mesons and baryons 
\cite{GURUHM2015,hmgururadius1}. With these set of parameters, we have estimated the electrical conductivity and the Hall conductivity using Eq.\eqref{equ19} and Eq.\eqref{equ20} as a function of temperature $(T)$ and baryon chemical potential ($\mu_B$) for different values of the magnetic field $(B)$.

\begin{figure}[!htp]
\begin{center}
\includegraphics[width=0.6\textwidth]{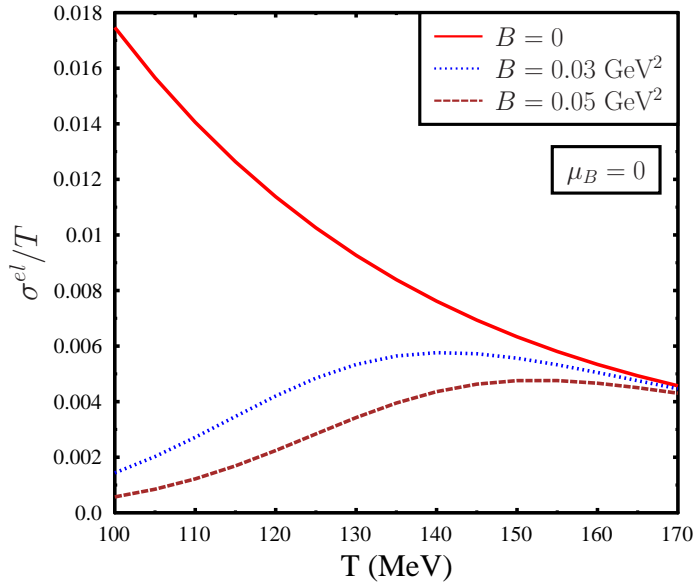}
\caption{Variation of normalized electrical conductivity ($\sigma^{el}/T$) with temperature ($T$) for different values of magnetic field ($B$) at zero baryon chemical potential. Red solid line represents $B=0$ case, blue dotted line and brown dashed line represents $B=0.03$ GeV$^2$ and $B=0.05$ GeV$^2$ respectively. With increasing magnetic field $\sigma^{el}/T$ decreases due to larger diffusion of charged particles transverse to the electric and magnetic field.}
\label{fig1}
\end{center}
\end{figure}

In Fig.\eqref{fig1} we show the variation of electrical conductivity $\sigma^{el}$ with temperature ($T$) for different values
of magnetic field ($B$) at vanishing baryon chemical potential ($\mu_B$). For vanishing magnetic field (B=0), the behavior of $\sigma^{el}/T$ is similar to the previous results, e.g. see Ref.\cite{MoritzGreif}. As may be observed from Fig.\eqref{fig1} $\sigma^{el}/T$ decreases monotonically with temperature at $B=0$. This can be associated with the increase of randomness of the system with larger collision rate leading to smaller relaxation time. We point out here that the dominant contribution to the electrical conductivity arises from the charged pions due to the small mass of the pions as compared to that of other hadrons. Thus the monotonic decrease of $\sigma^{el}/T$ is due to the decrease of relaxation time of pions with increasing 
temperature.

\begin{figure}[!htp]
\begin{center}
\includegraphics[width=0.6\textwidth]{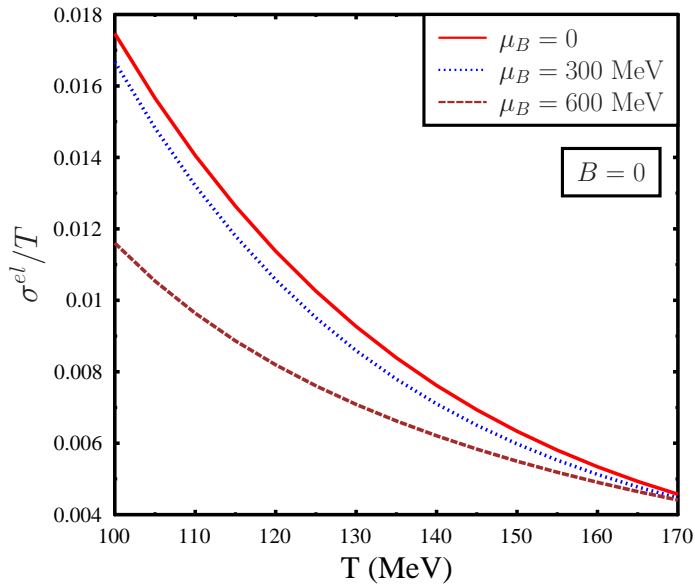}
\caption{Variation of normalized electrical conductivity $\sigma^{el}/T$ with temperature $T$ for different values of baryon chemical potential $\mu_B$ at zero magnetic field. Red solid line represents $\mu_B=0$ case. Blue dotted and brown dashed line represent electrical conductivity at finite baryon chemical potential with $\mu_B=300$ MeV and $\mu_B=600$ MeV respectively. In the case of zero magnetic field, with increasing $\mu_B$ normalized electrical conductivity $\sigma^{el}/T$ decreases predominantly due to decrease of the relaxation time of pions.}
\label{fig2}
\end{center}
\end{figure}

For non vanishing magnetic field, the behaviour of $\sigma^{el} /T$ is very different as compared to $B=0$ counterpart. Firstly, it is observed that with increase in magnetic field strength the electrical conductivity decreases. This decrease in electrical conductivity with the magnetic field can be understood physically. At zero magnetic field, the electric current is along the direction of the electric field. However,
at finite magnetic field, charges also diffuse transverse to both electric and magnetic field, due to the Lorentz force, giving rise to a reduced current along the direction of electric field.
This is also reflected in the expression for electrical conductivity as in Eq.\eqref{equ19}. The effect of magnetic field lies in the cyclotron frequency for the charged hadrons and occurs in the 
denominator in the Eq.\eqref{equ19}, leading to suppression of electrical conductivity. We would like to point out here, however, that for small magnetic field effects arising from Landau quantization for the charged hadrons can be neglected and hence 
the relaxation time arising from the scattering of the hadrons are independent of magnetic field. The total magnetic field dependence in the expression for electrical conductivity arises only from its dependence on the cyclotron frequency ($\omega_c$).

Secondly, the temperature dependence of $\sigma^{el}/T$ at finite magnetic field, as shown in the Fig.\eqref{fig1} is rather nontrivial with a non monotonic structure in contrast  to the monotonic decrease of its $B=0$ counterpart. For a non zero magnetic field $\sigma^{el}$ first increases with temperature and finally decreases. This behavior is mainly due to the combined effect of relaxation time and magnetic field, as can be seen from Eq.\eqref{equ19}. With increasing temperature thermal averaged relaxation time decreases. Hence the numerator of  Eq.\eqref{equ19}
decreases. But for non zero magnetic fields, $(1+(\omega_c\tau)^2)$ in the denominator also decreases with temperature. This interplay between $\tau$ and $\omega_c\tau$ decides the behavior of $\sigma^{el}$ with temperature. It is easy to see for large $\tau$ (low temperature) $\sigma^{el}\sim \frac{1}{\omega_c^2\tau}$, leading to an increasing behaviour of electrical conductivity with temperature. At higher temperature with small relaxation time $\sigma^{el}\sim \tau$ leading to decrease of electrical conductivity at higher temperature.
It is important however to note that due to the combination of $\omega_c\tau$ in Eq.\eqref{equ19}, mass of the particles explicitly enters in the electrical conductivity, apart from the same also in the distribution function. Hence at finite magnetic field contributions of the heavier particles may not be negligible.

Next we consider the variation of normalized electrical conductivity with temperature ($T$) for various values of baryon chemical potential ($\mu_B$) at zero magnetic fields is shown in the Fig.\eqref{fig2}. It is clear from this figure that with increasing chemical potential ($\mu_B$) electrical conductivity decreases. For the range of $\mu_B$ considered here the contribution to the electrical conductivity from the charged hadrons is dominated by the charged pions similar to the case with vanishing chemical potential. At finite chemical potential the pion relaxation time decreases with $\mu_B$ due to scattering with the baryons, mostly from the nucleons. One would have naively expected the nucleon contribution to the electrical conductivity to increase with $\mu_B$, which will lead to an increase in the total electrical conductivity due to the $\mu_B$ dependent distribution function in the expression of electrical conductivity. However this increase of the baryonic contribution to the electrical conductivity is not enough to compensate the decreasing contribution arising from pions, at least for the chemical potential considered in the present investigation. This leads to a decrease of the total electrical conductivity with increase in baryon chemical potential at vanishing magnetic field.

\begin{figure}[!htp]
\begin{center}
\includegraphics[width=0.6\textwidth]{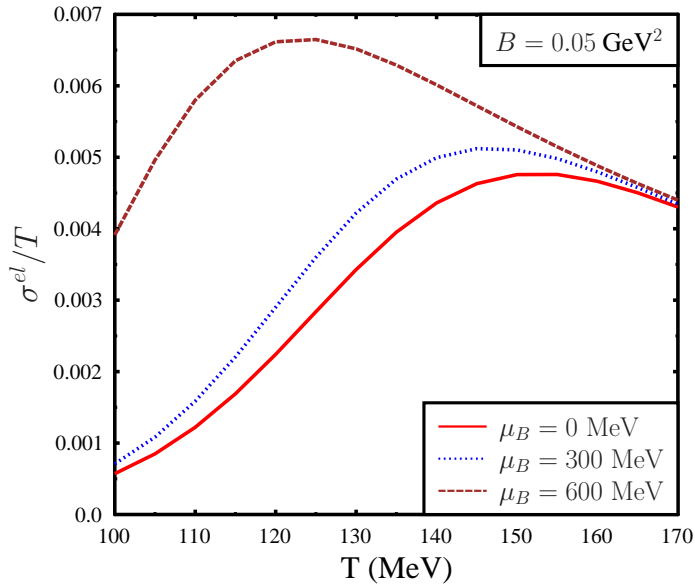}
\caption{Variation of normalized electrical conductivity $\sigma^{el}/T$ with temperature for different values of baryon chemical potential $\mu_B$ at $B=0.05$ GeV$^2$. Red solid line represents $\mu_B=0$ result. Blue dotted line and brown dashed line represents $\mu_B=300$ MeV and $\mu_B=600$ MeV results, respectively.}
\label{fig3}
\end{center}
\end{figure}

\begin{figure}[!htp]
\begin{center}
\includegraphics[width=0.48\textwidth]{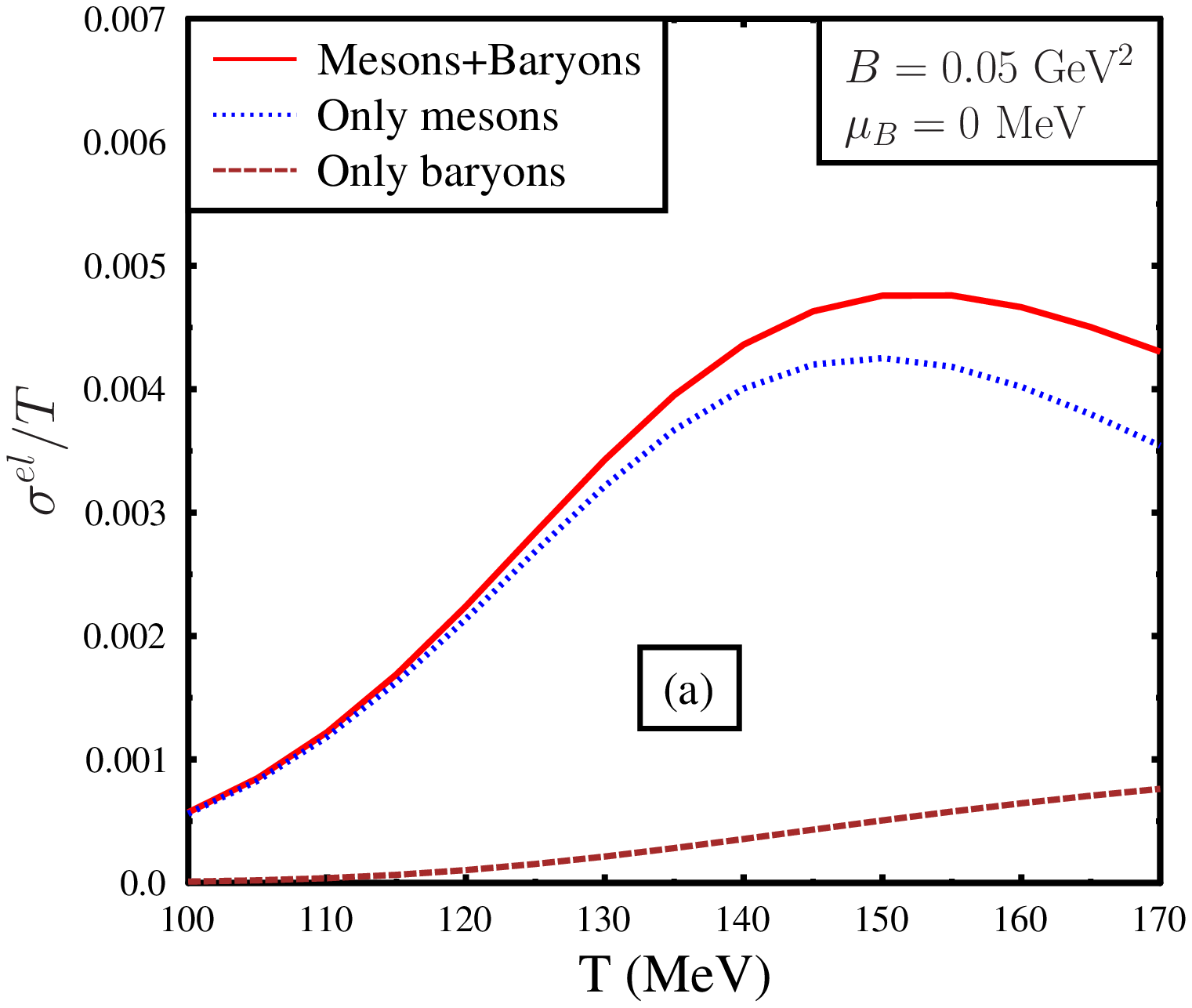}
\includegraphics[width=0.48\textwidth]{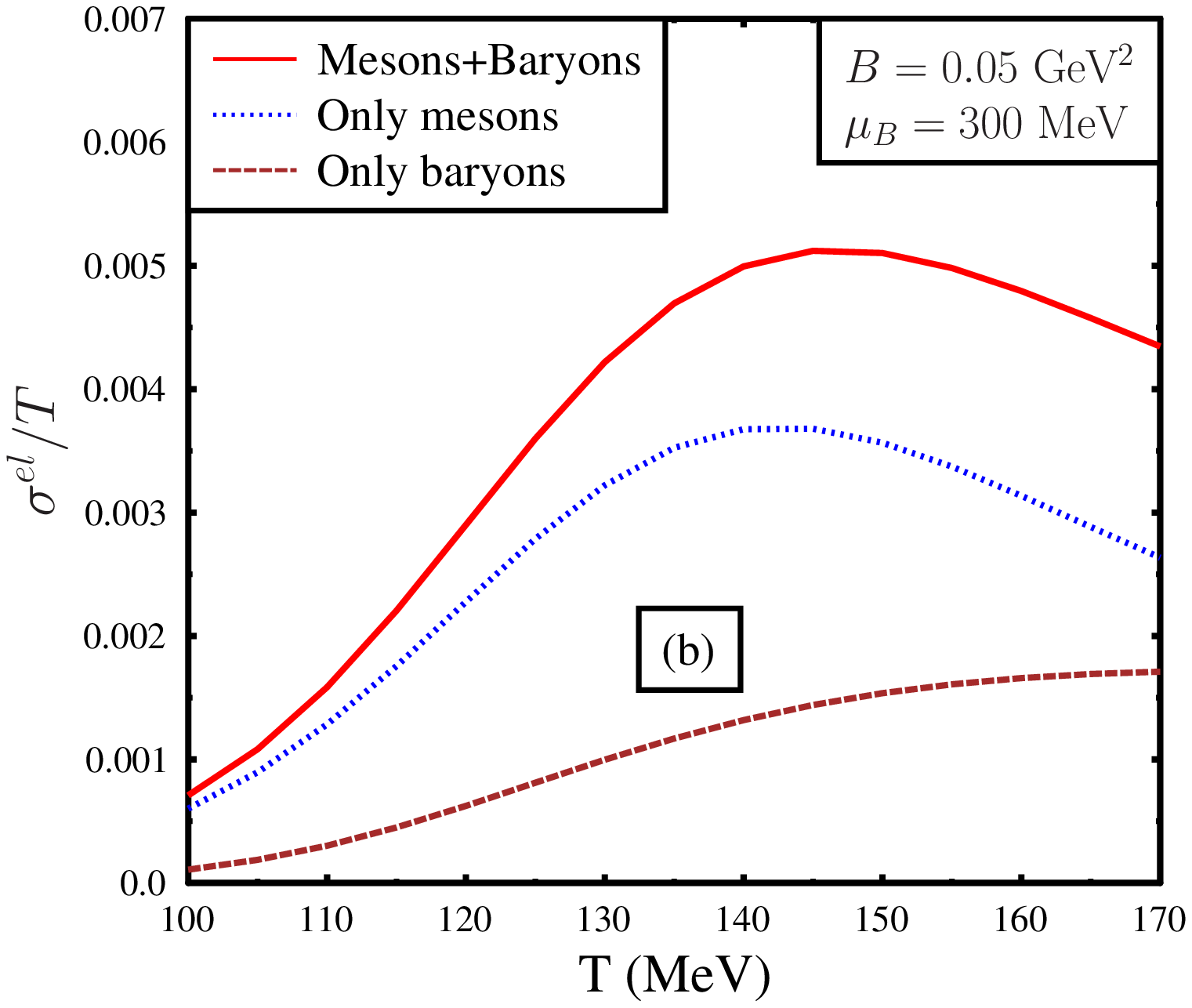}
\includegraphics[width=0.48\textwidth]{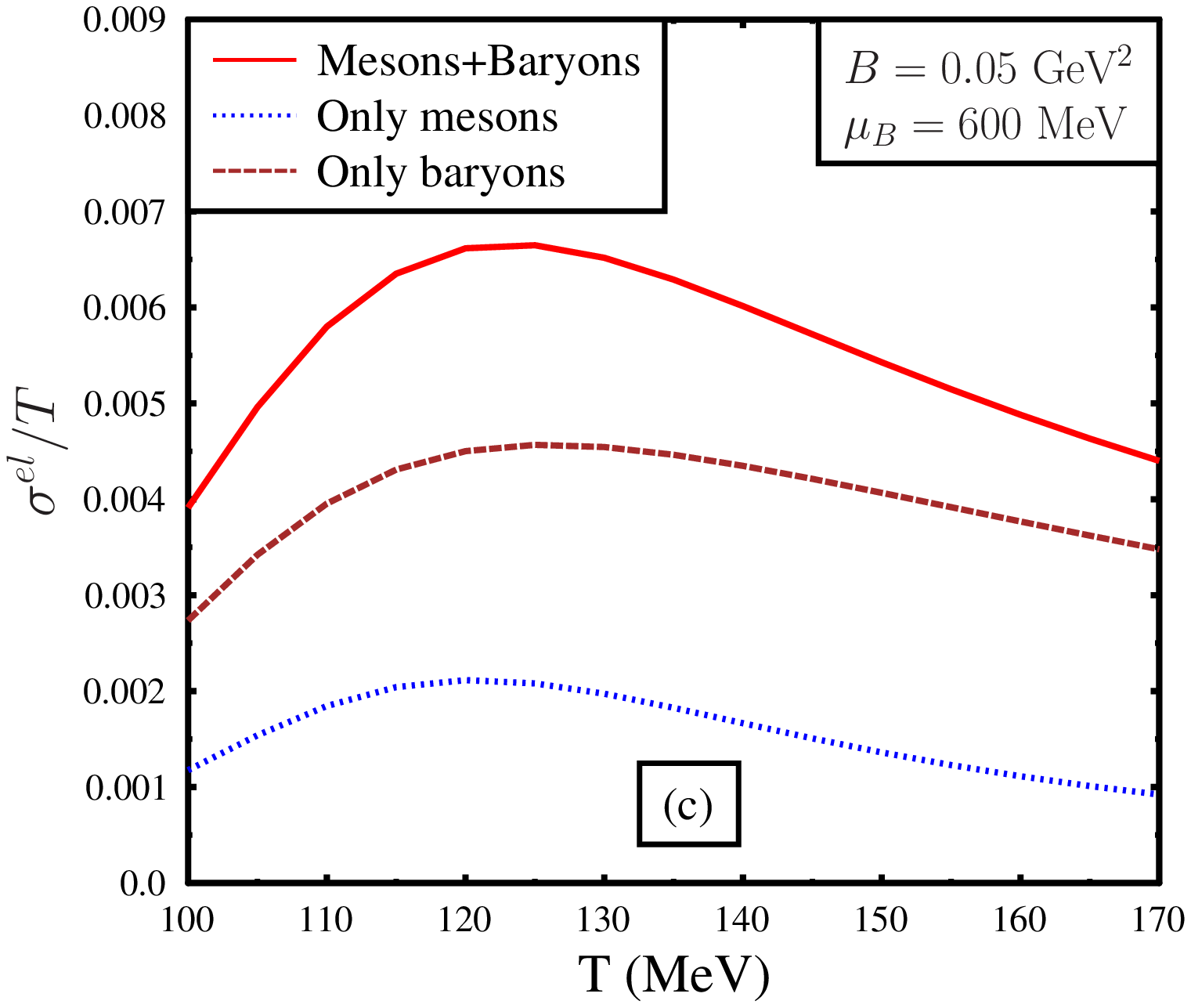}
\caption{Variation of normalized electrical conductivity $\sigma^{el}/T$ (red solid line) with temperature for different values of baryon chemical potential $\mu_B$ at $B=0.05$ GeV$^2$. Plot (a) is for $\mu_B=0$ MeV, plot (b) is for $\mu_B=300$ MeV and plot (c) is for $\mu_B=600$ MeV. Contribution of mesons (blue dotted ) and baryons (brown dashed) have been shown separately. With increasing baryon chemical potential baryon contribution to the total $\sigma^{el}$ increases and the mesonic contribution decreases.}
\label{fig4}
\end{center}
\end{figure}

Next we discuss the variation of normalized electrical conductivity $\sigma^{el}/T$ with temperature ($T$) in presence of magnetic field and for different values of baryon chemical potential ($\mu_B$). This is shown in Fig.\eqref{fig3}. Unlike the vanishing magnetic field case, it is seen that $\sigma^{el}/T$ increases with baryon chemical potential. This behaviour can be understood as follows. At finite magnetic field the contributions of the mesons to the electrical conductivity further decreases due to larger cyclotron frequency as compared to baryons, apart from the decrease in the relaxation time with increase in $\mu_B$. On the other hand the contributions of the baryons is enhanced at finite 
chemical potential. For the chemical potential considered here the contributions of the baryons become larger compared to that of mesons leading to an increase in the electrical conductivity with baryon chemical potential. This has been explicitly shown in Fig.\eqref{fig4}, where we have shown the contributions from baryons and mesons separately for $\mu_B=0,300$MeV and $600$ MeV respectively in \eqref{fig4}(a), \eqref{fig4}(b), \eqref{fig4}(c).

\begin{figure}[!htp]
\begin{center}
\includegraphics[width=0.6\textwidth]{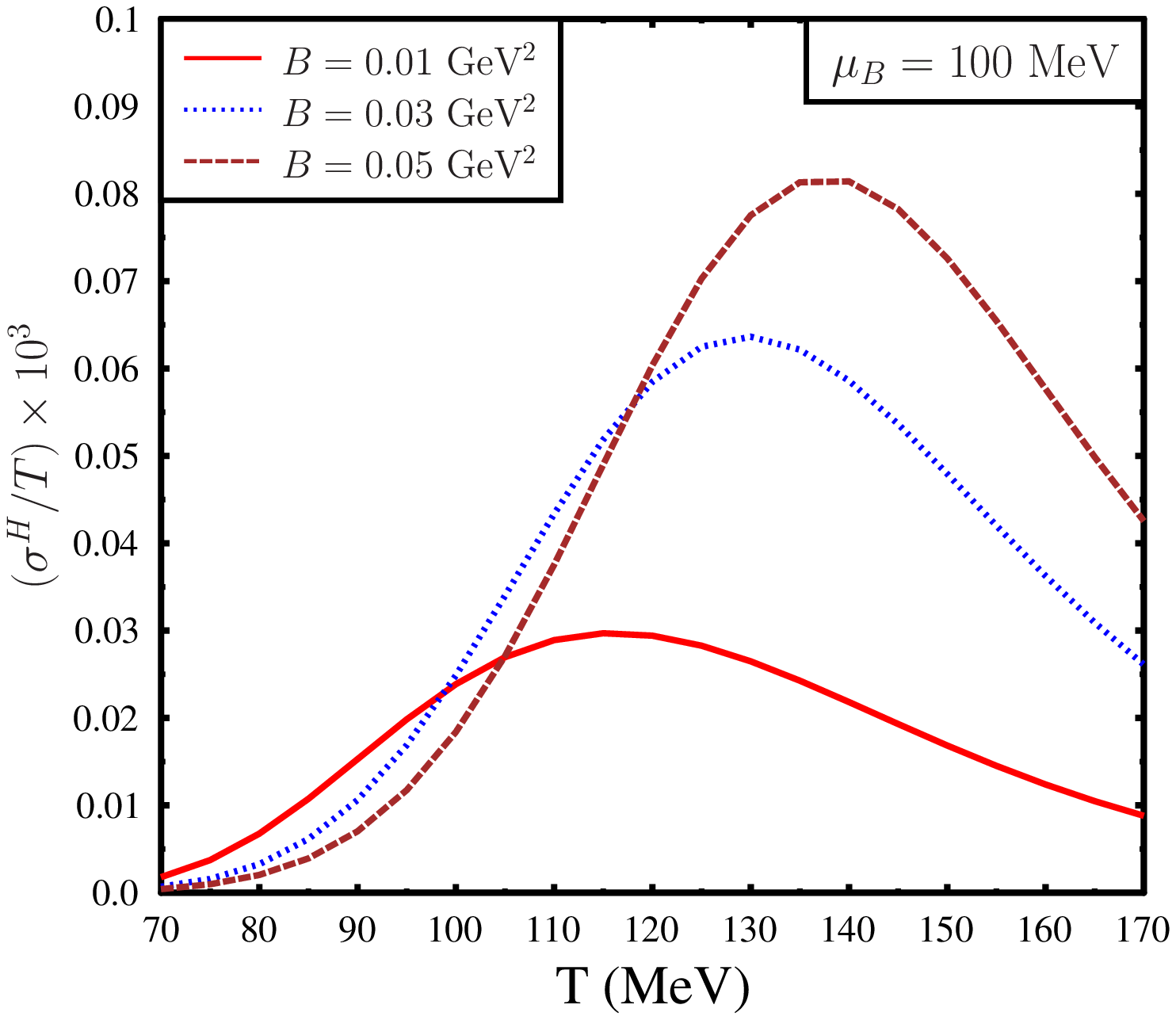}
\caption{Variation of normalized Hall conductivity $\sigma^H/T$ with temperature for various values of magnetic field at $\mu_B=100 MeV$. Red solid line, blue dotted line and the brown dashed line represent result for $B=0.01$ GeV$^2$, $B=0.03$ GeV$^2$ and $B=0.05$ GeV$^2$ respectively. With increasing magnetic field Hall conductivity increases, because for larger magnetic field more particles are deflected in a direction transverse to both electric and magnetic field.}
\label{fig5}
\end{center}
\end{figure}

Next, we discuss Hall conductivity in hadronic gas within HRG model. In Fig.\eqref{fig5}, we
show the variation of Hall conductivity with temperature $(T)$ for different 
values of the magnetic field at finite baryon chemical potential $\mu_B=100$ MeV. Let us note that 
due to the opposite gyration of the particles and the antiparticles in a magnetic field, the mesonic contribution to the Hall conductivity gets exactly canceled out. Hence, it is only the baryons which contribute to the Hall conductivity at finite baryon chemical potential. It may be observed in  Fig.\eqref{fig5} that for the small temperature the Hall conductivity decrease with increase in magnetic field, while for larger temperature the Hall conductivity increase with magnetic field. This can be understood from the expression of Hall conductivity in Eq.\eqref{equ20}. At low temperature since the relaxation time is smaller then the Hall conductivity the integrand $\sim \frac{1}{\omega_c\tau}$ ($\omega_c=\frac{eB}{\epsilon}$), which explains the suppression of Hall conductivity with increasing magnetic field.
On the other hand at large temperature with smaller relaxation time the integrand $\sim \omega_c\tau$ which explains the increase in the Hall conductivity with increasing magnetic field.
 Further, it is also interesting to observe that for a fixed value of the magnetic field, the normalized Hall conductivity first increases with temperature, reaching a maximum value and eventually decreases at a higher temperature. This behavior of the Hall conductivity with temperature is due to convolution of multiple factors e.g. the relaxation time, the magnetic field and the cyclotron frequency similar to the behaviour electrical conductivity for non vanishing magnetic field.

\begin{figure}[!htp]
\begin{center}
\includegraphics[width=0.6\textwidth]{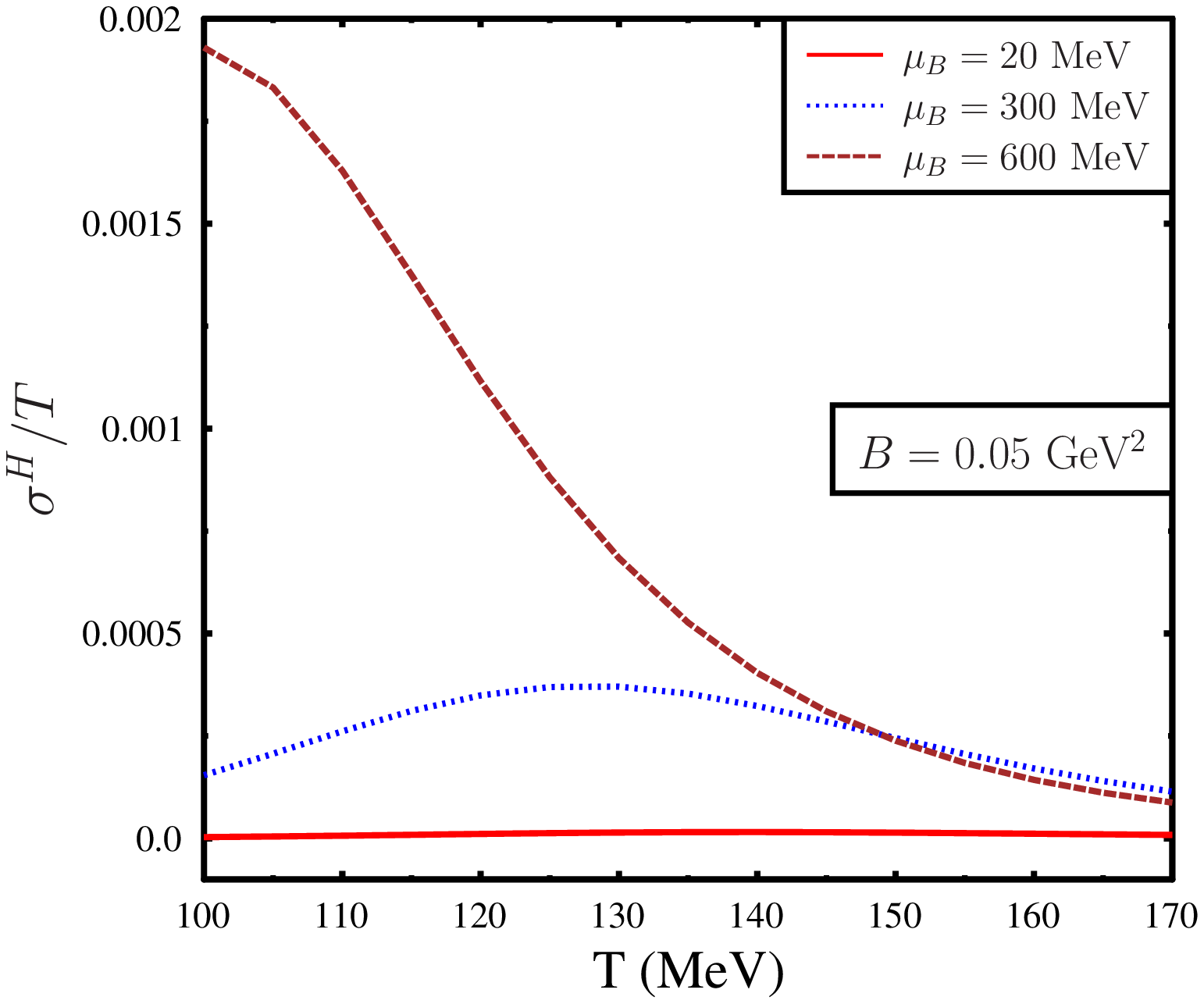}
\caption{Variation of normalized Hall conductivity $\sigma^H/T$ with temperature for various values of baryon chemical potential at $ B=0.05$GeV$^2$. Red solid line, blue dotted line and the brown dashed line represent result for $\mu_B=20$ MeV, $\mu_B=300$ MeV and $\mu_B=600$ MeV respectively. Due to large number density of baryons at finite baryon chemical potential, Hall conductivity increases with increasing baryon chemical potential.}
\label{fig6}
\end{center}
\end{figure}

In Fig.\eqref{fig6} we plotted the variation of the normalized Hall conductivity $\sigma^H/T$ with temperature for different values of baryon chemical potential at $B=0.05$ GeV$^2$. As may be noted from this figure for smaller chemical potential the Hall conductivity is smaller. This is due to the fact that for finite Hall conductivity the imbalance between the number of particles and antiparticles is required.  With increase in baryon chemical potential, the number density of particles are significantly larger than that of antiparticles leading to a  non vanishing Hall current. Again the non monotonic behavior of normalized Hall conductivity with temperature for a specific value of the magnetic field is similar to Fig\eqref{fig5}.

\begin{figure}[!htp]
\begin{center}
\includegraphics[width=0.48\textwidth]{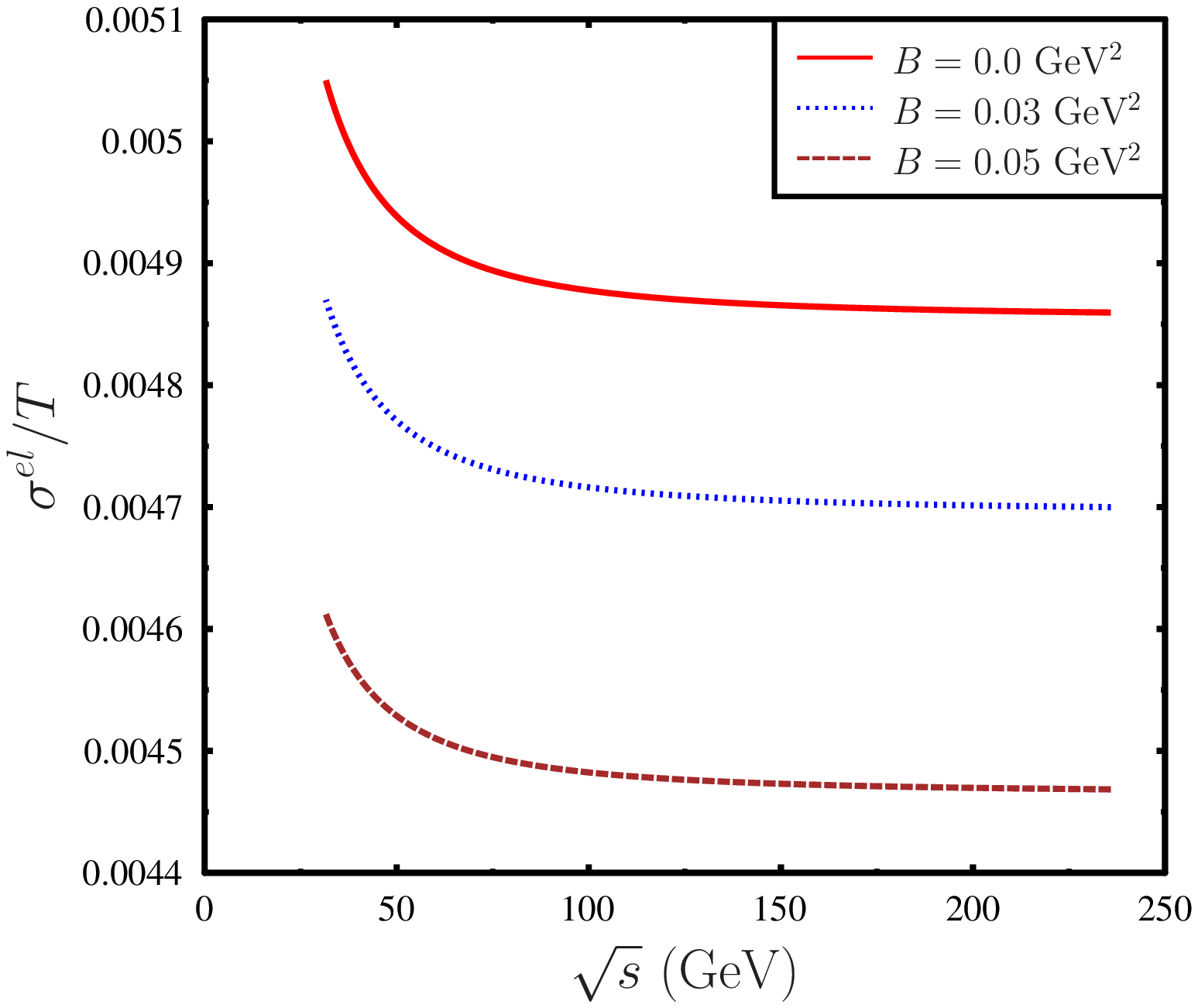}
\includegraphics[width=0.48\textwidth]{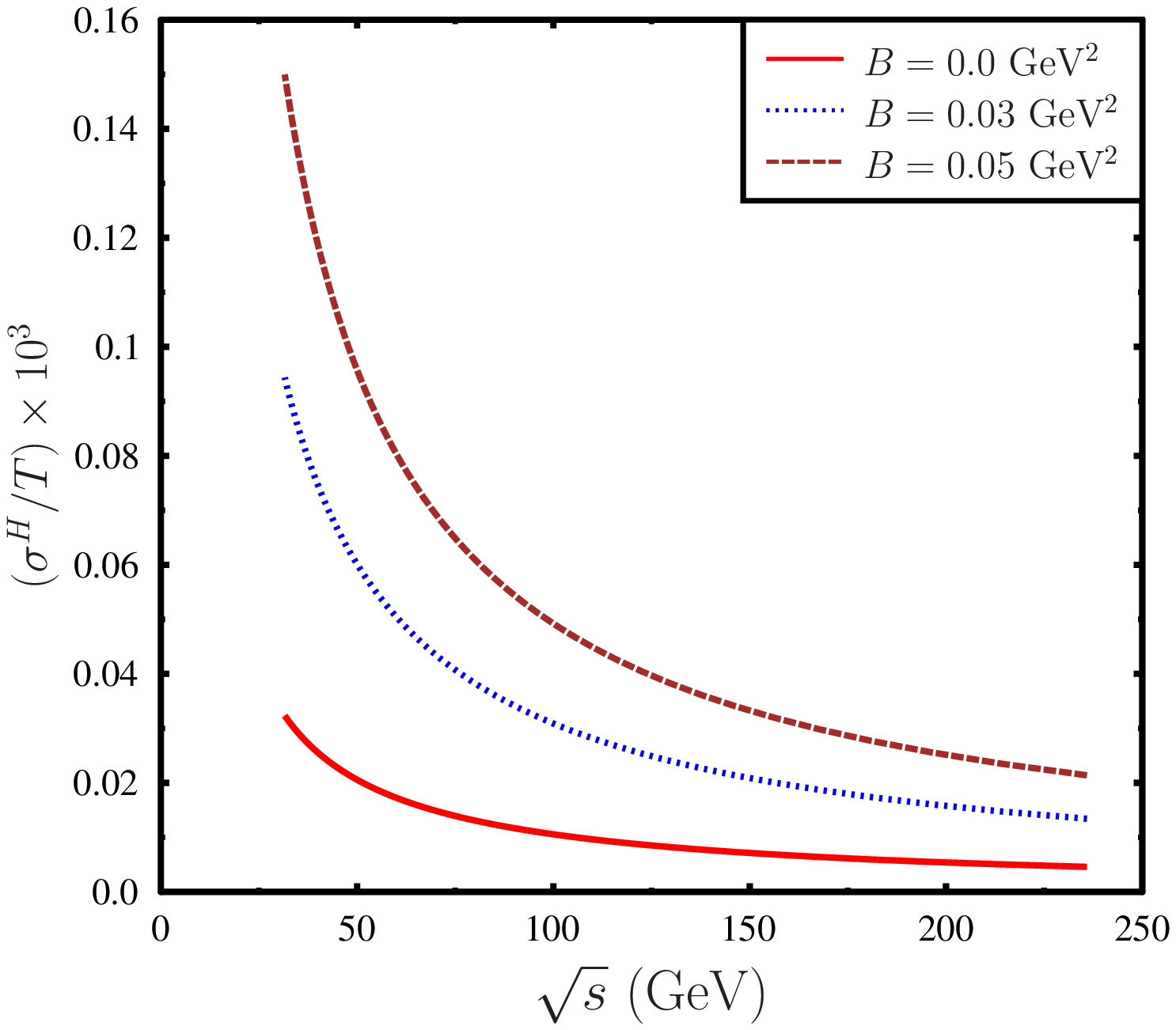}
\caption{Left plot: Variation of normalized electrical conductivity $\sigma^{el}/T$ with center of mass energy for difference values of magnetic fields, Right Plot: Variation of normalized Hall conductivity $\sigma^{H}/T$ with center of mass energy ($\sqrt{s}$) for different values of magnetic field. With increasing magnetic field $\sigma^{el}/T$ decreases and $\sigma^{H}/T$ increases.}
\label{fig7}
\end{center}
\end{figure}

Finally, in order to make a connection of our results with heavy ion collision experiments, we estimate the electrical conductivity and Hall conductivity as a function of beam energy $\sqrt{s}$. Fig.\eqref{fig7} shows the variation of $\sigma^{el}/T$ and $\sigma^{H}/T$ with center of mass energy $\sqrt{s}$ for different values of magnetic field. The dependence of $\sigma^{el}/T$ and $\sigma^{H}/T$ with the center of mass energy $\sqrt{s}$ are extracted using the statistical thermal model description  of the particle yield at various $\sqrt{s}$ \cite{freezeout1}. If we assume the freeze out parameters are independent of the magnetic field, then we could use the fitted freeze out temperature and baryon chemical potential as given by, $T(\mu_B)=a-b\mu_B^2-c\mu_B^4$, with $a=0.166\pm 0.002$ GeV, $b=0.139\pm0.016$ GeV$^{-1}$, $c=0.053\pm 0.021$ GeV$^{-3}$ and $\mu_B(\sqrt{s})=d/(1+e\sqrt{s})$, with $d=1.308\pm0.028$GeV, $e=0.273\pm0.008$ GeV$^{-1}$ respectively \cite{freezeout1}. In this investigation, we have considered the central values of the fitting parameters. Left plot and the right plot in Fig.\eqref{fig7} shows the variation of $\sigma^{el}/T$ and $\sigma^{H}/T$ with the center of mass energy $\sqrt{s}$ respectively, for various values of the magnetic field. From this figure, it is clear that with an increasing magnetic field $\sigma^{el}/T$ decreases and $\sigma^{H}/T$ increases. The variation of $\sigma^{el}/T$ and $\sigma^{H}/T$  with $\sqrt{s}$ for a given magnetic field is non trivial. At zero magnetic field $\sigma^{el}/T$ decreases with increasing temperature as well as increasing baryon chemical potential. Now with decreasing center of mass energy freeze out temperature decreases but freeze out baryon chemical potential increases. Hence one has to consider the effect of both temperature and baryon chemical potential to understand the variation of $\sigma^{el}/T$  with $\sqrt{s}$. Note that at finite magnetic field variation of $\sigma^{el}/T$ as well as $\sigma^{H}/T$ with temperature and chemical potential is more complicated than the case when the magnetic field is zero. Hence the variation of $\sigma^{el}/T$ and $\sigma^{H}/T$ along the freeze out curve is rather convoluted.

\section{conclusions}
In this investigation, we have estimated the electrical ($\sigma^{el}$) and the Hall conductivity ($\sigma^{H}$) of the hot and dense hadron gas in the presence of an external magnetic field. We have estimated the electrical and Hall conductivity by solving Boltzmann transport equation in the presence of the external electromagnetic field within the framework of relaxation time approximation. It is important to note that relaxation time approximation is valid when the external field is not strong. This assumption is valid in the case of hadron gas because at the time of freeze out strength of  the external magnetic field is relatively smaller than the initial magnetic field in heavy ion collisions. Also, in this case, we have not considered the Landau quantization of the charged particles as well as magnetic field dependent dispersion relation due to relatively smaller magnetic field. 

At vanishing magnetic field our results are similar to the previous results \cite{MoritzGreif} as shown in Fig.\eqref{fig1} and Fig.\eqref{fig2}. In the absence of magnetic field ($B=0$), pions are the dominant contributors in the electrical conductivity. With increasing temperature ($T$) and chemical potential ($\mu_B$) relaxation time of pion decreases. This gives rise to decreasing electrical conductivity with increasing temperature and baryon chemical potential.
However this situation is more complicated in the presence of magnetic field. With increasing magnetic field $\sigma^{el}$ decreases. This is because in the presence of magnetic field charged particles experience Lorentz force and they move in a direction transverse to the electric field and magnetic field. Since the baryons are heavier than the mesons, specifically pions, baryonic contributions can be important at non zero magnetic field. With increasing chemical potential more baryons contribute to the electrical conductivity. Hence with increased chemical potential electrical conductivity increases. Variation of electrical conductivity in the presence of a magnetic field with temperature is rather complicated. Variation of $\sigma^{el}$ with temperature is a convolution of various factors, e.g. temperature-dependent relaxation time, cyclotron frequency of different hadrons and also the distribution functions.
Generically with increasing temperature $\sigma^{el}$ first increases, becomes maximum at intermediate temperature and finally decreases. 

Unlike electrical conductivity $\sigma^{el}$, at zero baryon chemical potential Hall conductivity vanishes.  In a condensed matter system Hall conductivity can be non zero because in this case either electrons or holes are dominant charge carriers. However in case of pair plasma number density and mobility of particles and anti particles are same. Thus the Hall current due to particles and anti particles exactly cancel each other. This situation is similar to the case of hot hadron gas when baryon chemical potential $\mu_B$ is zero. Mesons do not contribute to the Hall conductivity. Only at finite baryon chemical potential, due to the difference in the numbers of baryons and anti baryons, one gets a non vanishing Hall conductivity. It is obvious that with increasing baryon chemical potential the Hall conductivity should increase due to the increase in the number density of baryons. This is exactly what we get as a numerical result shown in Fig.\eqref{fig6}. On the other hand the dependence of the Hall conductivity on magnetic field is non monotonic. While it increases with magnetic field for higher temperature, it decreases with magnetic field at lower temperature, which is evident from 
Fig.\eqref{fig5}. This is due to the dependence of the Hall conductivity on cyclotron frequency and relaxation time as given in Eq.\eqref{equ20}.
 We have already mentioned that at small baryon chemical potential Hall conductivity is small. This is the case at RHIC and LHC. However, for lower center of mass energy (e.g. FAIR and NICA) the chemical potential at freeze out is larger. In such situations the  Hall conductivity can be significant. 

\section*{Acknowledgement}
RKM would like to thank PRL for support and local hospitality for her visit, during which this problem was initiated. Also, RKM would like to thank Basanta K. Nandi and Sadhana Dash for constant support and encouragement.

\end{document}